\documentclass[prl,aps,twocolumn]{revtex4}

\usepackage{graphics}
\begin{document}

\title{NOISE SECURED INTERNET}
\author{{\textmd {\textbf }}\\
  \fontsize{11}{13}\selectfont
  Geraldo A. Barbosa\\
  \fontsize{9}{11}\selectfont
  \textit{QuantaSec, Consulting and Projects in Quantum
Cryptography Ltd.}\\
  \fontsize{9}{11}\selectfont
  \textit{Av. Portugal 1558, Belo Horizonte MG 31550-000
Brazil.}\\
  \fontsize{9}{11}\selectfont
  \textit{Email: GeraldoABarbosa@hotmail.com}\\[10pt]
}


\date{18 January 2006.}
\begin{abstract}
This work shows how a secure Internet can be implemented through a
fast key distribution system that uses {\em physical} noise to
protect the transmitted information. Starting from a shared random
sequence $K_0$ between two (or more) users, long sequences $R$ of
random bits can be shared and not involving a third party.  The
signals sent over the Internet are deterministic but have a built-in
Nature-made uncertainty that protects the shared sequences. After
privacy amplification the shared $R$ random bits --encrypted by
noise-- are subsequently utilized in one-time-pad data ciphering.
The physical generated protection is not susceptible to advances in
computation or mathematics.
In particular, it does not depend on the difficulty of factoring
prime numbers as many cryptography systems rely on.

\end{abstract}

\maketitle

\newcommand{\be}{\begin{equation}}
\newcommand{\ee}{\end{equation}}
\newcommand{\bea}{\begin{eqnarray}}
\newcommand{\eea}{\end{eqnarray}}

\section{Introduction}

The Internet is currently the main communication vehicle for citizens in general, banks and E-commerce. Protocols based on
mathematical complexities strive to offer a secure Internet while hackers attempt to break in for profits. As a matter of
fact, the existing Internet offers only tenuous security.  While a few security providers offer reasonable service within
the current technological landscape, they are vulnerable to technological advances. A search for new paradigms to establish
a secure Internet but not sensitive to technological or mathematical advances is ongoing.

This paper describes how to implement a practical secure communication system for the Internet while avoiding altogether
protocols based purely on mathematical complexities. This noise-encryption system relies on laws of Nature but also avoids
single-photon state protocols such as BB84 \cite{BB84}. Single-photon protocols cannot be amplified and therefore do not
work for the long-haul communications necessary for the Internet. Furthermore, signals from single-photon protocols cannot
be converted from optical to electrical and back to optical without loss of security. Nor they are practical for wavelength
multiplexing (WDM). These steps are necessary to the Internet. Alternative systems such as those using discrete or
continuous variable processes and relying on homodyne measurements (e.g., Ref. \cite{grangier}) are very sensitive to noise,
which leads to low key rate transfer, and cannot work in the naturally disturbed and complex Internet networks.

In the proposed implementation of a secure Internet deterministic ciphered signals go through arbitrary communication
channels.  They are ciphered by random signals from physical sources in nonorthogonal $M$-ry bases. This system has evolved
from a key distribution system recently proposed \cite{mykey,infoth}(See also \cite{alphaeta1} and \cite{alphaetaEXP}). This
secure Internet distributes deterministically random sequences of bits to be utilized in a fast ``one-time-pad'' scheme.

A simplified but imperfect illustration of this system based on physical noise would be a public radio station emitter A
that changes its carrier frequency in a very fast and truly random way. A user B who possesses  a perfect knowledge of this
random variation could set his tuner to automatically lock onto it. A clear sound or message would result. To an intruder
who does not possess any information on the carrier variations, only noise will be detected. Actually, the presented
cryptographic system does not rely on frequency variations but on random jumps among distinct nonorthogonal phase bases
where the bits are inscribed. This is as far as this analogy goes.

The security of the key distribution provided by this system relies on a few points: 1) A shared secrecy by A and B on a
starting key sequence $K_0$ and 2) a bit-by-bit uncertainty  Nature-made noise $N_i$ associated to each bit $R_i$ and
recorded  on a interleaved $M$-ry nonorthogonal basis. Knowledge of $K_0$ gives for the legitimate users the mapping of the
bases jumps in the emitter and thus the bit $R_i$ inscribed on each basis. Privacy amplification procedures statistically
exclude the eventually compromised fraction of shared bits. The sequences of random bits $R_i$ will be generated by a truly
random process and sent one-by-one between users A (Alice) and B (Bob). The batch of shared  secret bits $R_i$ will be used
subsequently in one-time-pad ciphering. The noise $N_i$ protects each bit $R_i$ from the attacker E (Eve) and provides the
information security level associated with all shared $R_i$.

While this  noise-secured Internet is logically equivalent to the optical system discussed on Refs. \cite{mykey} and
\cite{infoth} it has a major distinctive feature: In Ref. \cite{mykey} the noise arises as part of the signal measurement by
the attacker and is inherent to the optical field in a fiber channel.
Here the signal sent over the network is a recorded signal and, as such, it is deterministic but contains bit information
$R_i$ and the associated noise $N_i$. This is equivalent to recording the results of an exceedingly noisy experiment and
giving them to two researchers for interpretation: one that knows how to subtract the noise and the other one, the attacker,
that cannot get rid of the inherent noise.

\section{Basic scheme}

 The signals are created by a physical random generator
(PhRG). The noise $N_i$ associated with the bit $R_i$ inscribed onto the $M$-ry nonorthogonal basis ($M\geq 2$) produces the
uncertainty measured by the attacker. This implies that the the emitter has to be equipped to detect and record the signals
generated by the PhRG. In other words, the definition of the measuring system is made by the emitter, not the attacker. The
signal sent is the signal controlled and measured by the emitter with a detection system of his choice. No restrictions are
placed on the attacker to obtain the exchanged signals on a public channel. She may obtain perfect copies of the
transmission. The signals emitted by the legitimate user obey constraints imposed to provide full security. Among the
advantages of the proposed system are: 1) Any public channel may be used for transmission (optical fibers, TV, microwave,
and so on); 2) The deterministic signals can be amplified with no security loss; 3) Signals can be converted from
electromagnetic to electrical and back to electromagnetic with no security loss; 4) Wavelength multiplexing is allowed on
the network; 5) Current Network and IP protocols can be used with no modifications for users in any IP classes.
\begin{figure}
\centerline{\scalebox{0.47}{\includegraphics{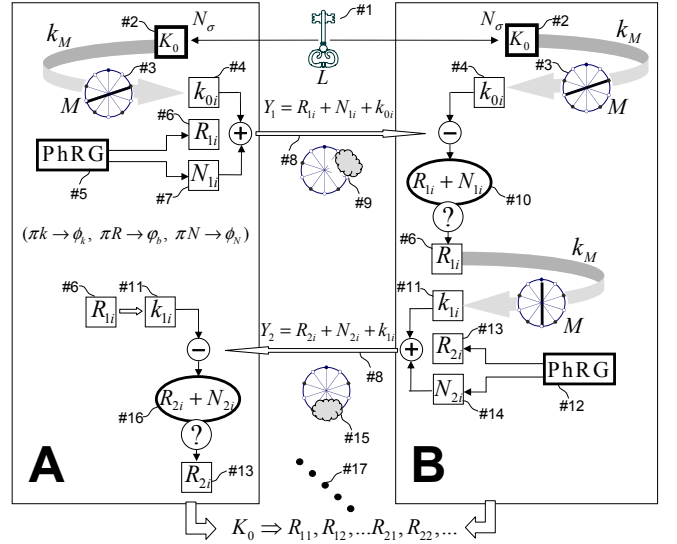}}}
\caption{A sketch of one cycle of operations of the key distribution
process in the Noise Secured Internet is shown. }
\label{InternetBlock}
\end{figure}
Fig. \ref{InternetBlock} shows a block diagram for one cycle of the
key distribution system. It describes how legitimate users A and B
distribute or transfer fresh random key bits generated by a PhRG (to
be described ahead). Just to be more general, a description starting
 with a $M-$ry system of levels will be presented. At the end, this system will be simplified
 for a speed-up in the communication process with no security loss.

 A and B share a
starting random key sequence (\#1) designated by $K_0$ (\#2) of
length $L$ (See Fig. \ref{InternetBlock}). These $L$ bits  are
divided into blocks of size $k_M$
 ($b(k_M), b(k_{M-1}), ... b(k_1)$) and each block defines randomly a
 basis $k_{0i}$ over a nonorthogonal set of bases
 on a ciphering wheel (\#3)  with $M$
bases, where $M=2^{k_M}$.
\begin{eqnarray}
\label{basis}
 k_{0i}=b(k_M) 2^{k_{M-1}}+b(k_{M-1})
2^{k_{M-2}}+...b(k_1) 2^0\:\:.
\end{eqnarray}
Given a $k_{0i}$ value (\#4), a bit 0 could be inscribed, e.g., in a uniform ciphering wheel such as the one shown in Ref.
\cite{mykey} where the phase values defining each basis are given by
\begin{eqnarray}
\label{wheel}
 \phi_{k_{0i}}=\pi \left[
\frac{k_{0i}}{M}+\frac{1-(-1)^{k_{0i}}}{2} \right],
\:k_{0i}=0,1,...M-1,
\end{eqnarray}
and a bit 1 will be inscribed displaced by $\pi$ with respect to bit 0 over each basis. A PhRG (\#5) generates random bits
$R_{1i}$ (\#6) that A would like to transfer securely to B. These signals contain noise $N_{1i}$ (\#7) with a natural phase
distribution (e.g., Gaussian distributed) of width $\sigma_{\phi}$. $R_{1i}$ can be understood in phase units (rd): values 0
or $\pi$ for bits 0 and 1. The Gaussian distribution width $\sigma_{\phi}$ (Set such that $\sigma_{\phi}< \pi/2$) may be
written $\sigma_{\phi}=\pi N_{\sigma_{\phi}}/M$ where $N_{\sigma}$ is the number of bases covered by $N_i$ (See Ref.
\cite{mykey}). The signal to be sent over the generic Internet communication channel (\#8) (network and servers) is
$Y_1=R_{1i}+N_{1i}+k_{0i}$. The combined effects of $N_{1i}+k_{0i}$ is to hide the bit value $R_{1i}$ on the ciphering wheel
(\#9). Although containing random information $Y_1$ is a deterministic signal and as such can be amplified and converted
into different signals through arbitrary nodes without any loss of security.

B has to extract $R_{1i}$ from $Y_1$. To this end he utilizes the same sequences from $K_0$ utilized by A to generate the
base values $k_{0i}$ (\#4). He subtracts this value from $Y_1$ and obtains $R_{1i}+N_{1i}$ (\#10) and obtain signals in {\em
binary} bases (single $k_i$ value). The effect of the noise $N_{1i}$ on Bob's {\em binary} basis is negligible because
$\sigma_{\phi}< \pi/2$ and his decision on the bit value is easy; therefore, he obtains $R_{1i}$ (\#6). From the received
sequence $R_i$ he forms bit blocks of length $k_M$ and constructs a new base sequence $k_{1i}$. The next steps are similar
to the first ones. Bob's PhRG (\#12) generates signal containing bits $R_{2i}$ (\#13) associated to noise $N_{2i}$ (\#14).
The signal $Y_2=R_{2i}+N_{2i}+k_{1i}$ is sent over the communication channel (\#8). The bit value $R_{2i}$  is hidden by the
overall noise $N_{2i}+k_{1i}$ (\#15). From her knowledge of $R_{1i}$ (\#6)  and, therefore, $k_{1i}$ (\#14), Alice subtracts
$k_{1i}$ from $Y_2$ and obtains $R_{2i}+N_{2i}$ (\#16). On her {\em binary} basis she easily obtains $R_2i$ (\#13). The
first cycle is complete. A and B continue to exchange random sequences as in the first cycle. The shared sequences
($R_{1i},...,R_{2i},...$), after a privacy amplification process, are the random bits to be subsequently utilized for
one-time-pad cipher.

Note that while for noiseless signals $Y_1=b$ and $Y_2=b$ carrying a repeated bit $b$, one has $Y_1\oplus Y_1=0$, noisy
signals give $Y_1=b+N_1$ and $Y_1=b+N_2$ and, therefore, $Y_1\oplus Y_1=N_1+N_2(=0\:\mbox{or}1)$. This frustrates several
correlation attacks.

\section{Some security considerations}

Security analyses were presented for the purely optical counterparts of this $M$-ry key distribution system for Internet and
are equally valid here. Ref. \cite{mykey} presented a bit-by-bit analysis and \cite{mykeyquantph} showed that the attacker's
initial uncertainty on the whole sequence of shared bits is equal to her uncertainty of the starting shared sequence $K_0$.
This dependence on the first key shared key sequence $K_0$ can be seen from the mutual information, defined by
\begin{eqnarray}
I(R:Y(R))=H(R)-H(R|Y(R))\:\:.
\end{eqnarray}
$I(R:Y(R))$ is used to write the difference between the mutual information between B and E in one cycle of length $L$:
\begin{eqnarray}
\Delta I&=&I_B-I_E\nonumber\\&=&\left[H(R)-H(R|Y_B(R))\right]-\left[H(R)-H(R|Y_E(R))\right]\nonumber\\
&=&H(R|Y_E(R))-H(R|Y_B(R))\:\:.
\end{eqnarray}
Eve's uncertainty on $R$ given $Y_E(R)$ is maximal ($H(R|Y_E(R))\rightarrow L$) while Bob may obtain the whole sequence $R$
from $Y_B$: $H(R|Y_B(R))=0$. Therefore, $\Delta I \leq L$ in the first cycle. Applying the chain rule
\begin{eqnarray}
H(L_1,L_2,...,L_n|Y_E)=\sum_{i=1}^n H(L_i|Y,L_1,...,L_{i-1})\:\:,
\end{eqnarray}
one may see that
\begin{eqnarray}
&&H(L_1,L_2,...,L_n|Y_E)=\hspace{5cm}\nonumber\\&&=\!\!H(L_1|Y_E)+H(L_2|Y_E,L_1)+H(L_3|Y_E,L_1,L_2)..
\end{eqnarray}
{\em If} the starting key ($K_0$ with length $L=L_1$) is open to
Eve, she obtains $L_2$ ($H(L_2|Y_E,L_1)=0$) in the same way as Bob
and also obtains all keys in the subsequent rounds. The noise level
superposed to the bits sent are designed to hide each $L_j$ from the
attacker. Privacy amplification --a necessary step-- applied to the
shared bits discards information eventually leaked to the attacker
and defines the final shared length of secure bits.

\section{The physical random generator}

\begin{figure}
\centerline{\scalebox{0.37}{\includegraphics{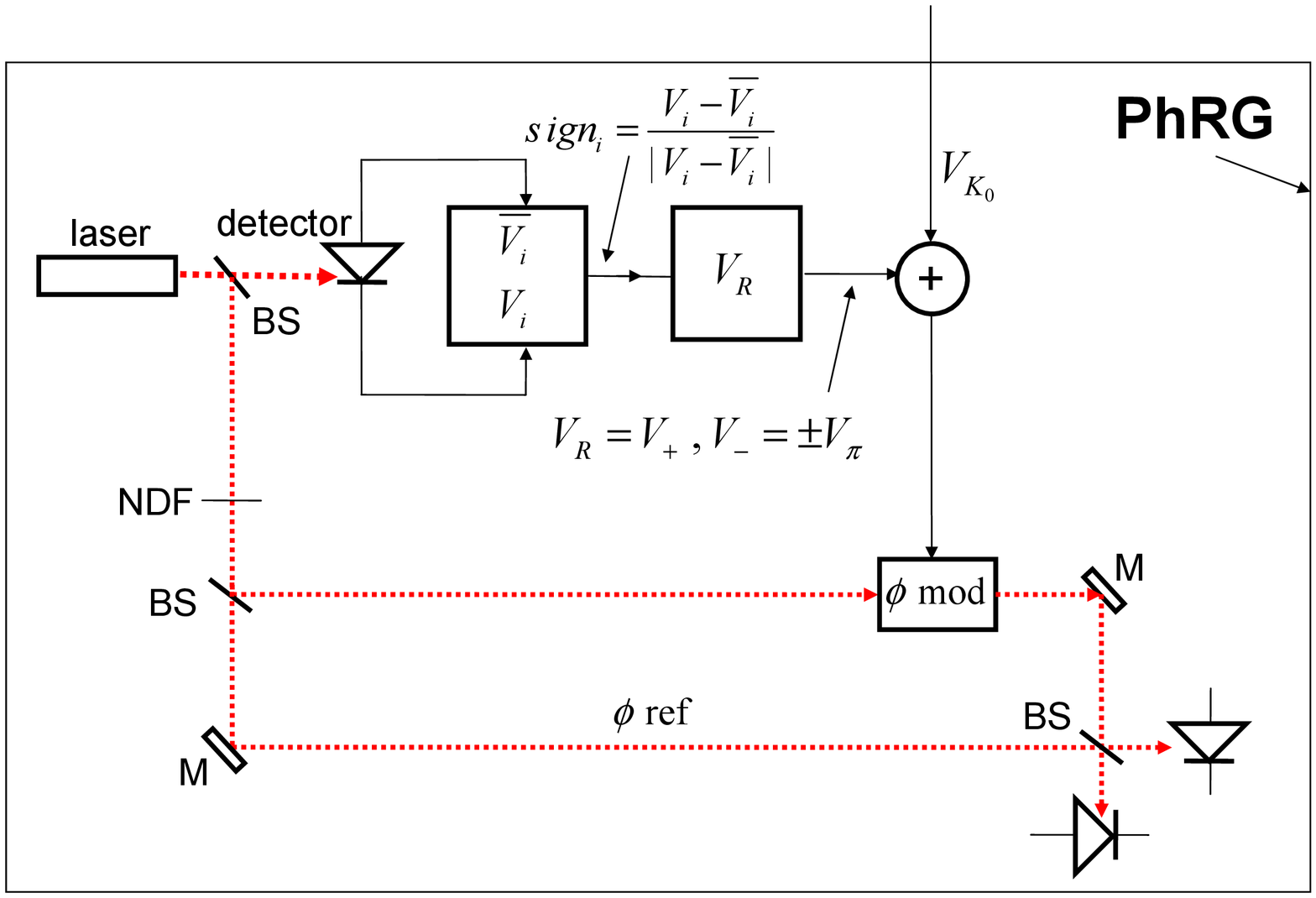}}} \caption{Sketch of PhRG with a coherent light source. This modulus
can be added externally or externally to a computer. The laser beam is divided by a beam splitter BS. The upper part shows a
detecting system where signals $V_i$ are generated corresponding to the sign of the generated signal with respect to the
average signal intensity. These binary signals are converted into binary voltages $V_R=\pm V_{\pi}$ that constitute fresh
random bits to be shared by A and B. The bottom part shows an interferometer with an optical phase modulator ($\phi$ mod) in
one of the arms. The laser beam is adjusted to an adequate intensity by a neutral density filter (or automatized filter).
Voltage values $V_{K_0}$ defining $M$-ry phase bases (e.g, $M=2$) are added to $V_R$ and applied to the phase modulator.
Detectors at the interferometer output produce the phase signals carrying basis, bit and noise information shown in Fig.
\ref{InternetBlock} as $Y_i$.} \label{PhRG}
\end{figure}
Fig. \ref{PhRG} sketches the PhRG and the input $V_{K_0}$ containing recorded bases information. The PhRG generate voltage
signals corresponding to bits $V_R$. These signals are added to the basis information  supplied by $V_{K_0}$ and  supplied
to a phase modulator in one arm of an optical interferometer. Output light is detected and converted to phase signals (with
respect to the laser field) that also contain phase noise associated with coherent light. These signals are written $Y_i$ in
Fig. \ref{InternetBlock}. As shown in Refs. \cite{mykey} and \cite{infoth}, this phase modulation of coherent signals
produce signals that carry a phase uncertainty given by the Gaussian distribution
\begin{eqnarray}
p_u \simeq e^{- (\Delta \phi)^2 /2 \sigma_{ \phi}^2}\:\:,
\end{eqnarray}
where $\sigma_{\phi}=\sqrt{2/\langle n \rangle}$ and $\langle n \rangle$ is the average number of photons in one bit.
  While several design variations are possible, Fig. \ref{PhRG} shows
basic parts to be considered. For secure transmission of signals the physical randomness is necessary as no known
mathematical algorithm has been proven to generate true random numbers. Several physical sources may be used alternatively
such as optical or thermal sources. However, optical sources can be much faster than the thermal ones and are therefore
necessary when speed is required. A PhRG can be seen as modulus that can be hooked (internally or externally) to a computer
linked to the Internet either in dedicated use or open to users such as in a cybercafe. In such a public system users may
generate and record on portable memories a batch of secure keys or use them to exchange one-time-pad ciphered information.

 One may also wonder about the cost of a brute force attack to determine the starting key $K_0$ from the transmitted
signals. Under the assumption that the uncertainty presented to the attacker covers $N_{\sigma}$ bases, the attacker would
know that the basis $k_i$ used in a given transmission is around a given region within the uncertainty $N_{\sigma}$. For a
$M$-ry system of uniformly spaced bases this amounts that only a set of less relevant bits $b(k_{\sigma})$ in Eq.
(\ref{basis}) hide the correct basis.  These $b(k_{\sigma})$ bits could be permutated in $b(k_{\sigma})!$ ways. As each bit
could be either 0 or 1 the total number of permutations to be searched for each bit emission would be $(\log_2 N_{\sigma})!
N_{\sigma}$. For the total number of $K_0$ bits the number of combinations would be
\begin{eqnarray}
C=2^{K_0}(\log_2 N_{\sigma})! N_{\sigma}\:\:.
\end{eqnarray}
Under this example of a uniform ciphering wheel exemplified by Eq. (\ref{wheel}), it is understood that the attacker may
know the fraction $1-(N_{\sigma}/M)$ of the total number of shared bits $k_M$ used by A and B to cipher a fresh generated
bit. For a sequence of L shared bits, Eve may obtain $L [1-(N_{\sigma}/M)]$ bits among $L$ because they were not covered by
noise. These bits have to be subsequently discarded by A and B through privacy amplification processes.

\section{Simplified bases}

\begin{figure}
\centerline{\scalebox{0.35}{\includegraphics{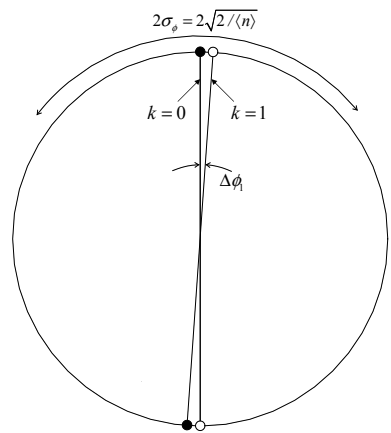}}} \caption{A ciphering set of bases in a phase sector with $M=2$.
$\sigma_{\phi}$ is the standard deviation in the phase caused by fluctuations in the light field. $\Delta \phi_1$ is the
spacing between two bases and should be kept  $\Delta \phi_1\ll \pi/2$.  $\langle n \rangle$ is adjusted so that $\pi/2
> \sigma_{\phi}\gg \Delta \phi_1$, e.g. $\langle n \rangle< 100$. Two states or bits can be inscribed on each basis. Dark
circles indicate positions for a bit 0 and open circles give possible positions for a bit 1.} \label{M2Sector}
\end{figure}
Use of a non-uniform set of bases leads to a more economical system: instead of a uniformly spaced circle of phases given by
Eq. (\ref{wheel}) one may use just a sector of phase values where the number of bases is just $M=2$. See Fig.
\ref{M2Sector}. The sector width or bases separation is made less than $2 N_{\sigma}$. Therefore, all bases will be within
the phase fluctuations caused by the noise $N_j$.  Phase positions on this sector are given by
\begin{eqnarray}
 \phi_{k_{0i}}= \left[\! k_{0i} \Delta \phi_1+ \pi \frac{1-(-1)^{k_{0i}}}{2} \right] , \:\:k_{0i}=0,1.
\end{eqnarray}
 With this sector of phase bases the
number of possible combinations for a brute force attack searching for all possibilities that may lead to $K_0$ is $C=2
\times 2^{K_0}\:\:.$
Reasonable $K_0$ lengths could be, say, $\sim 10^6, 10^9$; they give a number of combinations $C$ to be tried that is not
computationally feasible.

Ref. \cite{infoth} derived explicit equations for the mutual information of the process and showed numerical examples to
quantify security in terms of the difference of the mutual information functions for A and B and A and E ($I_{AE}=\epsilon
I_{AB} \:,\: \epsilon \ll 1$). As has also been shown, the security of one sequence sent depends on the secrecy of the
former sequence received. This is an a-priori condition over which $I_{AB}>I_{AE}$ follows. Statistically, the attacker may
acquire some bits correctly and the legitimate users have to use privacy amplification protocols to eliminate that possible
amount of information acquired by Eve. Privacy amplification randomly reduce the number of bits to eliminate possible
information leaked to Eve. A and B are able to share a large number of random sequences $R \gg K_0$ before the bit
reconciliation and privacy amplification steps severely shorten the length $L$ of the sequences. These length reductions
lead to a slow down of the process and eventually to its halt. A convenient minimum length $L_{min}$ can be chosen so that a
new fresh sequence $K_0$ restarts the whole process. See Ref. \cite{infoth} for a discussion on the distillation process.
The amount of possible leakage can be estimated using mutual information functions as shown in Ref. \cite{infoth}.
 It was shown that a fraction $1-f$ ($f\sim 0.9991$ in the example given) could be compromised in every cycle; therefore,
$\sim 0.1$\% should be eliminated by privacy amplification protocols. As a result, the remaining fraction $f$ of bits in
every cycle is secure. A and B then succeed after many cycles in sharing a number of secure bits much larger than the
initial shared sequence $K_0$. With the example given in Ref. \cite{infoth}, after sharing $\sim 10^3 L$  bits, $\sim 660 L$
(66\%) will be distilled by privacy amplification.
Renewal of fresh starting sequences $K_0$ was discussed in Refs. \cite{mykey} and
\cite{infoth}. Ancient methods such as hand-to-hand delivery and steganography could be used for some applications. Even
certified key providers may be acceptable for some uses. The slow BB84 key distribution process could also be used to
distribute the starting sequences $K_0$ with proven security; the speed of the key distribution process will then boosted by
the Noise Secured Internet system described here.

\section{Conclusions}

It has been shown that two (or more) Internet users starting from a shared secret sequence of random bits $K_0$ and adding a
simple ``hardware'' modulus (PhRG) to their computers will succeed in generating a large number of secret keys to be used in
one-time-pad cipher. The system works at optical speed and does not require any special Internet protocol. Signals
associated with noise are generated in the PhRG and the signals to be sent are deterministic ones. The associated security
is not related to protocols based on mathematical complexities in current use. This system is proposed as a possible new
paradigm for a secure Internet.

\noindent Email: geraldoabarbosa@hotmail.com\\
Phone: Brazil(31)3441-4121


\end{document}